\documentclass[final,5p,times,twocolumn]{elsarticle}

\usepackage{graphicx}
\usepackage{amsmath}

\renewcommand{\Vec}[1]{{\bf #1}}

\journal{Physica E}

\begin{document}

\begin{frontmatter}



\title{Optical Hall conductivity in 2DEG and graphene QHE systems}


\author[label1]{Takahiro Morimoto}
\author[label2]{Yasuhiro Hatsugai}
\author[label1]{Hideo Aoki}

\address[label1]{Department of Physics, University of Tokyo, Tokyo, Japan}
\address[label2]{Institute of Physics, University of Tsukuba, Tsukuba, Japan}

\begin{abstract}
We have revealed from a numerical study that 
the Hall plateaus are retained in 
the optical Hall conductivity $\sigma_{xy}(\omega)$ 
in the ac ($\sim$ THz) regime 
in both of the ordinary two-dimensional electron gas and graphene 
in the quantum Hall regime, 
although the plateau height  in ac deviates from integer multiples of 
$e^2/h$.  
The effect remains unexpectedly robust 
against a significant strength of disorder, which we attribute 
to an effect of localization.  
We predict the ac Hall plateaus are observable through the 
Faraday rotation with the rotation angle characterized by the fine-structure constant $\alpha$.
In this paper we clarify the relationship between plateau
structures and the disorder strength by performing numerical calculation.
\end{abstract}


\end{frontmatter}


\section{Introduction} 
While the quantum Hall effect (QHE) enjoys a long history over 
almost three decades, QHE physics still harbors 
avenues that have not been fully explored.  
Here we wish to bring to light that 
the {\it optical Hall effect} is indeed a new avenue, 
which does harbor an unexpected plateau structures.   
One motivation for the study 
is that recent experimental spectroscopy 
in the THz regime are becoming so advanced 
that optical measurements  (e.g., Faraday rotation in magnetic fields) 
begin to be feasible 
for QHE systems 
in the THz regime .\cite{hangyo,ikebe2008cds}  
Theoretically, we pose here a fundamental question of 
how QHE,  a topological phenomenon \cite{tknn,hatsugai1993cna}, evolves into the {\it optical Hall conductivity}, especially in the THz 
($\sim$ cyclotron energy) regime. 

Another motivation is the recent emergence of physics of 
graphene with its ``massless Dirac particle",
whose anomalous QHE has been experimentally observed\cite{Nov05,Kim-gr}.  For graphene QHE system,
optical properties begin to be studies, among which 
are experimental transmission spectra\cite{sadowski}, or 
theoretical examination of the cyclotron emission\cite{morimoto-CE}.
One interest about graphene is  the $n=0$ Landau level (LL) in graphene QHE, which should be unusual since the level consists of a mixture of electrons and holes around the Dirac point. Special features of $n=0$ LL are intensively discussed, including its robustness against long-ranged random magnetic fields or corrugations of graphene.\cite{Nov05,nomura2008qhe,guinea-ripple,kawarabayashi09}

Motivated by these, we have theoretically calculated the optical Hall conductivity $\sigma_{xy}(\omega)$ in the quantum Hall regime 
with the Kubo formula 
for both the ordinary 2DEG and graphene.   The effect of long-ranged potential disorder is taken into account with the exact diagonalization method
to incorporate the effects of localization due to disorders,
since the ac 
conductivity,  even for 2DEG,  
has only been dealed with a phenomenological 
(Drude) formalism\cite{oconnell82} or with Maxwell's equations\cite{peng1991frq}.
We shall unravel from the numerical study the following: 
(i) The plateaus in $\sigma_{xy}(\omega)$ in the optical (THz) region, 
although not quantized, 
are unexpectedly robust against disorder up to a significant degree of disorder.  
We attribute the unexpected robustness to an effect of localization.
(ii) For graphene,  the optical Hall plateaus are again retained in the THz regime, while the cyclotron resonance structure reflects the graphene Landau 
levels that are not uniformly spaced due to the `massless Dirac' band.  
(iii)
 Experimentally the step structure in $\sigma_{xy}(\omega)$ should be measured 
as steps in the Faraday rotation, for which 
we can estimate $\Theta_H \simeq
\alpha \simeq 7 \ \mbox{mrad}$, 
which is well within the experimental resolution \cite{ikebe2008cds}
and may be called the fine-structure constant $\alpha$ seen as a rotation, 
while $\alpha$ has been visualized from transmission in \cite{nair-alpha}.
In this paper we present the details of our calculation
and clarify the relationship between
plateau structures and the disorder strength by extensive numerical calculation,
which has not been done in our related work\cite{morimoto-opthall}.

\section{Optical Hall conductivity in 2DEG} 
Let us first look at the optical Hall conductivity 
in the QHE in 2DEG 
as realized in GaAs/AlGaAs with a Hamiltonian,
$H_0=\frac{1}{2 m^*}(\Vec{p} +e \Vec{A})^2$.  
To include the effect of disorder, we employ the exact diagonalization method for  disorder 
described by randomly placed scatterers with a potential 
$$
V(\Vec r)=\sum_j u_j \exp(-|\Vec r-\Vec R_j|^2/2d^2)/(2\pi d^2),
$$
where $d$ is the range of the potential, while the strength of the potential, 
each placed at $\Vec R_j$, is assumed to take $u_j=\pm u$ with  
random signs so that the density of states broadens symmetrically in energy.  
We adopt $d=0.7\ell$, which is comparable to  the magnetic length $\ell=\sqrt{\hbar/eB}$.  
The degree of disorder can be characterized by 
$\Gamma = 2u [n_{\rm imp}/2\pi(\ell^2+d^2)]^{1/2}$, which 
is a measure of the Landau level broadening\cite{ando,morimoto-CE} 
with $n_{\rm imp}$ being the density of impurities.

We have diagonalized the Hamiltonian $H_0+V$ 
by representing $V$ with the basis of the free Hamiltonian $H_0$,
$
\psi_{n,k_y}=\left[2^n n! \sqrt{\pi}\ell \right]^{-1/2}\exp\left[-\frac{1}{2}(\frac{x-\ell^2 k_y}{\ell})^2\right] H_n(\frac{x-\ell^2 k_y}{\ell})\frac{1}{L}e^{-i k_y y},
$
where $H_n$ is the Hermite polynomial and $L$ the system size.

To calculate the optical conductivity in the QHE system we 
use the Kubo formula, 
\begin{equation}
\begin{split}
\sigma_{\alpha\beta}(\omega) 
= \frac{e^2\hbar}{i\pi}\int d\varepsilon \frac{f(\varepsilon)}{\hbar\omega}
&\left[\mbox{Tr}\left( j_{\alpha}{\rm Im}G(\varepsilon)j_{\beta}(G^+(\varepsilon+\hbar \omega)-G^+(\varepsilon))\right) \right. \\
&\hspace{-0.5em} \left. -\mbox{Tr}\left( j_{\alpha}(G^-(\varepsilon)-G^-(\varepsilon-\hbar \omega))j_{\beta}{\rm Im}G(\varepsilon)\right)\right],
\end{split}
\end{equation}
where $\alpha,\beta=x,y$, $f(\varepsilon)$ the Fermi distribution, and 
$G^{\pm}\equiv G(\epsilon \pm i \delta)$ with a positive infinitesimal $\delta$.
At $T=0$ which we assume here, the formula is reduced to
\begin{eqnarray}
\sigma_{xy}(\omega)
&=&
\frac{i\hbar e^2}{ L^2}
\sum_{\epsilon_a < \varepsilon_F}\sum_{\epsilon_b \ge \varepsilon_F}
\frac 1 {\epsilon_b-\epsilon_a}
\nonumber\\
&\times&
\left(\frac{j_x^{ab}j_y^{ba}}{\epsilon_b-\epsilon_a-\hbar\omega} 
-\frac{j_y^{ab}j_x^{ba}}{\epsilon_b-\epsilon_a+\hbar\omega}
\right) ,
\label{kuboformula}
\end{eqnarray}
where $\epsilon_a$ is the eigenenergy, and $\varepsilon_F$ the Fermi energy.
The current matrix 
elements, $j_x^{ab}$, are calculated through the current matrices between 
the free Landau levels,
\begin{eqnarray}
j_{x}^{n,n'}=
i\sqrt{\hbar \omega_c/2m^*}\left(\sqrt{n}\delta_{n-1,n'}-\sqrt{n+1}\delta_{n+1,n'}\right),\nonumber\\
j_{y} ^{n,n'}=\sqrt{\hbar \omega_c/2m^*}\left(\sqrt{n}\delta_{n-1,n'}+\sqrt{n+1}\delta_{n+1,n'}\right),
\end{eqnarray}
where $n, n'$ are the Landau indices and $\omega_c$ the cyclotron frequency.  
Here we have retained 7 Landau levels in the basis, 
which is sufficient in the energy range considered here 
for $L= 15\ell$.  
For the ensemble average
we have taken 5000 random configurations for the impurity potential.

\begin{figure}[t]
\includegraphics[width=\linewidth]{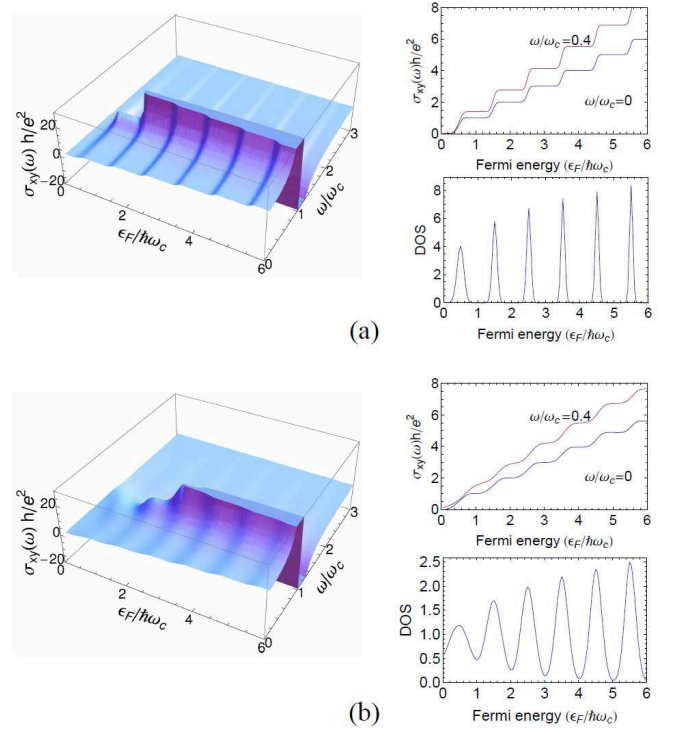}
\caption{
Exact diagonalization results for
the optical Hall conductivity $\sigma_{xy}(\varepsilon_F,\omega)$ (left 
panels), 
the optical Hall conductivity $\sigma_{xy}(\varepsilon_F,\omega =0.4\omega_c)$ as 
compared with the static one, and
the density of states (bottom right) 
for the usual QHE system in the presence of long-range impurities with the impurity strength $\Gamma/\hbar\omega_c=$0.2(a), 0.7(b).
}
\label{longrange-gaas}
\end{figure}

Figure \ref{longrange-gaas} shows the results for 2DEG.  
We plot 
$\sigma_{xy}$ on an $(\omega, \varepsilon_F)$ plane, 
where the $\omega=0$ cross section corresponds to the 
familiar static Hall conductivity, which is quantized into $ne^2/h$.  
We immediately notice two features: 
(i) $\sigma_{xy}(\omega)$ for a fixed value of $\varepsilon_F$ 
exhibits a resonance structure around 
the cyclotron frequency (as observed in the experiment\cite{hangyo}).  
(ii) Away from the resonance, 
if we look $\sigma_{xy}(\omega)$ for a fixed frequency,
{\it a step-like structure} is preserved  as a function of the Fermi energy $\varepsilon_F$.  
Although the step heights are not quantized exactly,
the flatness is surprisingly retained as seen in 
Fig.\ref{longrange-gaas}. 
(iii) The density of states (DOS) shown in Fig.\ref{longrange-gaas} broadens with the disorder strength, where the levels having smaller 
Landau indices tend to be destroyed faster.  By contrast, 
ac as well as static Hall conductivity $\sigma_{xy}(\varepsilon_F)$ in Fig.\ref{longrange-gaas}
 are blurred with $\Gamma$ much more slowly and the Hall step structure remains.
As compared with the self-consistent Born approximation\cite{morimoto-LT-opthall}, with which the broadening of DOS and the width of the plateau-to-plateau transition are solely determined by $\Gamma$, 
the present exact diagonalization approach includes the localization effect and the associated plateau structure in $\sigma_{xy}(\varepsilon_F)$.

The step structure is in fact a quantum effect (outside 
the Drude picture).  In the dc QHE, the localization is the cause of the 
plateaus in the Aoki-Ando picture\cite{aoki-ando1981}. 
%\bibit{aoki-ando1981} H.  Aoki  and T. Ando, Solid State Commun. {\bf 38}, 1079 (1981).
In the ac QHE, the Kubo formula, eqn(1), contains $\omega$, 
and does not simply reduce to a topological expression.
In this sense the result for the robust plateaus is quite nontrivial. 

The physical insight for the unexpectedly robust ac Hall step structure
is that the main contribution to the optical Hall conductivity comes
from the extended states 
whose existence ensures the robust step structure in the ac Hall conductivity.
To be more precise, the magnitude of the current matrix elements in eqn(\ref{kuboformula}) is much larger for the extended states than for localized 
states, so that the optical Hall conductivity is dominated by the transitions 
between the extended states 
which reside around the center of each Landau level, while 
the localized states give rise to the step structure. 
Namely, 
if we consider that the current matrices are significant 
only between extended states 
in the Landau levels just below and above $\varepsilon_F$\cite{aoki78}, 
we can collect the contributions 
from the extended states with the energy denominator, 
$\varepsilon_{b}-\varepsilon_a$, replaced 
with $\hbar\omega_c$ in 
eqn(\ref{kuboformula}) to reduce the equation to
\begin{equation}
\begin{split}
\sigma_{xy}(\omega)
&\simeq
\frac{e^2\hbar}{i L^2}
\sum_{\varepsilon_a<\varepsilon_F \le \varepsilon_b}^{\rm extended}
\frac 1 {(\hbar\omega_c)^2}
\\
&\times
\left(j_x^{a,b}j_y^{b,a}\frac{\omega_c}{\omega_c-\omega}
-j_y^{a,b}j_x^{b,a}\frac{\omega_c}{\omega_c+\omega}\right).
\end{split}
\end{equation}
Then $\sigma_{xy}(\omega)$ should exhibit a step structure 
every time $\varepsilon_F$ traverses the energy region for 
the extended states with the frequency dependence 
close to $\omega_c^2/(\omega_c^2-\omega^2)$ resonance structure.

\section{Optical Hall conductivity in graphene} 
We now turn to the optical Hall conductivity in graphene.
Here we again adopt the exact diagonalization method for 
the disorder potential introduced by randomly placed scatterers.  
When the range of the random potential 
is much larger than the lattice constant in graphene, the 
scattering between K and K' points in the Brillouin zone is suppressed, 
so that we can assume the random term takes a 
diagonal form in the Dirac Hamiltonian as
\begin{equation}
H_0+V=v_F 
\begin{pmatrix}
0 & \pi^- &0&0 \\
\pi^+ &0&0&0 \\
 0&0&0 & \pi^+ \\
 0&0& \pi^- &0 \\ 
\end{pmatrix}
+
\begin{pmatrix}
1&0&0&0\\
0&1&0&0\\
0&0&1&0\\
0&0&0&1\\ 
\end{pmatrix}
V(\Vec r)
.
\label{longrange-ham}
\end{equation}
So we adopt the Dirac model as in Refs.\cite{nomura-ryu-koshino,nomura2008qhe} 
to obtain wave functions and conductivity in the presence of disorder.
After diagonalizing the Hamiltonian $H_0+V$ by retaining 9 Landau levels with the system size $L=15\ell$, 
we have obtained eigenstates in the presence of disorder in the same manner as in the 2DEG QHE system.
Then we have calculated $\sigma_{xy}(\varepsilon_F,\omega)$ with the Kubo formula eqn(\ref{kuboformula}) with the current matrix 
for graphene\cite{ZhengAndo,morimoto-CE}:
\begin{eqnarray}
j^{x} _{n,n'}=\frac{v_F}{\hbar}C_{n} C_{n'}\left[ {\rm sgn}(n)\delta_{|n|-1,|n'|}+{\rm sgn}(n')\delta_{|n|+1,|n'|}\right],
\nonumber\\
j^{y} _{n,n'}=i \frac{v_F}{\hbar} C_{n} C_{n'}\left[ {\rm sgn}(n)\delta_{|n|-1,|n'|}-{\rm sgn}(n')\delta_{|n|+1,|n'|}\right]
\label{matrixelement}
\end{eqnarray}
with $C_n= 1 (n=0)$ or $1/\sqrt{2}$ (otherwise),
which has a selection rule, $|n|\leftrightarrow |n+1|$, 
peculiar to graphene.

\begin{figure}[t]
\includegraphics[width=\linewidth]{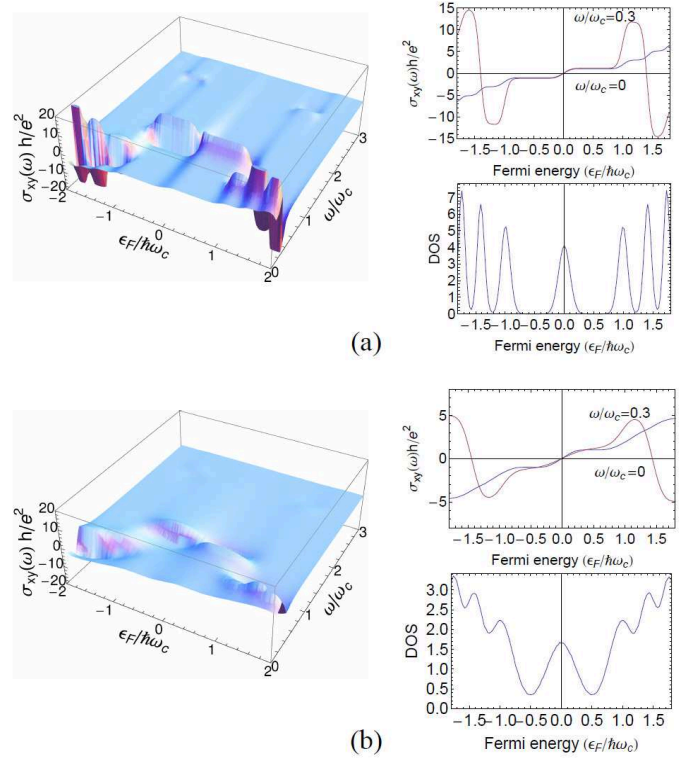}
\caption{
Exact diagonalization results for
the optical Hall conductivity $\sigma_{xy}(\varepsilon_F,\omega)$ (left panels), 
the static and optical Hall conductivities, and
the density of states (bottom right)
for the graphene QHE system in the presence of long-range impurities with the impurity strength $\Gamma/\hbar\omega_c$=$0.2(a), 0.5(b)$.
}
\label{longrange-gr}
\end{figure}

In the result, Fig.\ref{longrange-gr}, 
we notice several features distinct from the 
result for the ordinary QHE system (Fig.\ref{longrange-gaas}):  

(i) The optical Hall conductivity 
$\sigma_{xy}(\varepsilon_F,\omega)$ exhibits a more complex 
structure, which reflects the Landau levels, 
${\rm sgn}(n)\sqrt{|n|}\hbar \omega_c$ with $\omega_c=v_F\sqrt{2eB/\hbar}$, 
that are not uniformly spaced for the massless Dirac dispersion.  
Thus a series of resonances appear around 
many allowed transitions, $|n|-|n'|=\pm1$.  

(ii) Away from these resonances, we again observe that 
{\it step-like structures} remain in the optical Hall conductivity, 
as clearly seen in Fig.\ref{longrange-gr}.  
Due to the electron-hole symmetry, $\sigma_{xy}(\varepsilon_F, \omega)$ is 
odd in $\varepsilon_F$ throughout, so the step structure is symmetric as well.

When we more precisely look at the result, 
DOS in Fig.\ref{longrange-gr} broadens with the disorder strength, where 
the nonuniform level spacing causes $n\neq 0$ Landau levels merge faster than $n=0$ Landau level does, and a trace of $n=0$ Landau level is visible even for $\Gamma=0.7\hbar\omega_c$.
In the static Hall conductivity $\sigma_{xy}(\varepsilon_F)$ , $n\neq 0$ Hall steps are smeared out as soon as Landau levels are merged, while the $n=0$ Landau level robustly remains in all the cases studied, 
which tells us that the extended states in the $n=0$ Landau level is robust while those in $n\neq 0$ Landau levels are not.
This confirms the specialty of the $n=0$ Landau level.
In the optical Hall conductivity $\sigma_{xy}(\varepsilon_F,\omega)$, larger-$n$ Hall steps are destroyed around smaller values of $\omega$ because of denser cyclotron resonances $|n|\leftrightarrow |n+1|$.

Especially, the $n=0$ step remains not only for $\Gamma=0.2\hbar\omega_c$ (Fig.\ref{longrange-gr}(a)) but also for $\Gamma=0.5\hbar\omega_c$ (Fig.\ref{longrange-gr}(b)), which is a significant disorder as seen in the merged DOS.  The step is visible until 
the cyclotron resonance at $\omega=\omega_c$ is approached, 
although Hall plateaus are somewhat blurred in the ac region.
In the static Hall conductivity $\sigma_{xy}(\omega=0)$,
 $n\neq 0$ Hall steps are smeared as soon as Landau levels are merged 
while the step associated with $n=0$ Landau level 
is relatively robust, 
which indicates that the extended states in $n=0$ Landau level is 
unusually robust.\cite{nomura2008qhe,kawarabayashi09}  
The present ac result indicates that 
the step in $\sigma_{xy}(\varepsilon_F,\omega)$ associated with $n=0$ Landau level exhibits robustness against disorder in the ac regime as well, 
which we take to be the effect of localization, 
while the plateau-to-plateau transition is related to the 
persistent delocalized states which, for $n=0$, cannot float due to 
the electron-hole symmetry.

\section{Faraday rotation} 
Finally let us mention the experimental feasibility.  
We propose that the ac Hall steps should be 
observable through accurate Faraday-rotation measurements in the THz to 
far-infrared spectroscopy.  
The Faraday-rotation angle $\Theta_H$ is directly connected to 
the optical Hall conductivity via
\begin{eqnarray}
\Theta_H &=&
\frac{1}{2}\arg \left(\frac{t_+(\omega)}{t_-(\omega)}\right)
\nonumber\\
&=&\frac 12 \mbox{arg} \left[ \frac{n_0+n_s+(\sigma_{xx}+i\sigma_{xy})/(c\epsilon_0)}{n_0+n_s+(\sigma_{xx}-i\sigma_{xy})/(c\epsilon_0)} \right]
\nonumber\\
&\sim&
\frac{1}{(n_0+n_s)c\varepsilon_0}\sigma_{xy}(\omega),
\end{eqnarray}
where $t_\pm$ is the transmission coefficients for circularly polarized light,
$c$ is the velocity of the light,
$n_0 (n_s)$ is the refractive index of air (substrate), 
and we have assumed $n_0+n_s\gg\sigma_{\pm}/(c\varepsilon_0)$ on 
the last line. 
In the QHE regime, the Faraday rotation angle is proportional 
to $\sigma_{xy}(\omega)$, 
so that the step structure in $\sigma_{xy}(\omega)$ should be observed 
as jumps in Faraday-rotation measurements.  
We can estimate the size of the 
jumps $\Delta\Theta_H$ by putting $\sigma_{xy}\sim e^2/h$ (when 
$\omega$ is well below the resonance), so that
\begin{eqnarray}
\Delta\Theta_H
\sim
\frac{1}{(n_0+n_s)c\varepsilon_0}\frac{e^2}{h}
\sim
\frac{2}{n_0+n_s}\alpha
\quad\sim
7 \ \mbox{mrad},
\end{eqnarray}
where $\alpha=e^2/(2\varepsilon_0 h c)$ is the fine-structure constant. 
So the steps in the Faraday-rotation angle should be 
of the order of the fine structure constant.  Recently Shimano {\it et al.} have achieved an experimental resolution of $\sim 1$ mrad 
\cite{ikebe2008cds}, so that the present 
effect is well within the experimental feasibility.  
Nair et al.\cite{nair-alpha} have seen the fine-structure constant 
from visual transparency of graphene, and the present 
proposal amounts to the fine-structure constant seen from a rotation.  

To summarize, we have revealed that the  
optical Hall conductivity in both the usual 2DEG QHE and graphene QHE systems 
has plateau structures that persist even in ac regimes 
for significant strengths of disorder. 
We wish to thank Ryo Shimano, Yohei Ikebe, Seigo Tarucha, Takeo Kato and Takashi Oka for 
illuminating discussions.  
TM was supported by Grant-in-Aid for the Japan Society for
Promotion of Science (JSPS) Research Fellows.
This work has been supported in part by Grants-in-Aid for Scientific Research, 
No.20340098 (YH, HA), 20654034 (YH) from JSPS and 
Nos.220029004, 20046002 (YH) from MEXT.




\begin{thebibliography}{22}


\expandafter\ifx\csname natexlab\endcsname\relax\def\natexlab#1{#1}\fi
\expandafter\ifx\csname bibnamefont\endcsname\relax
  \def\bibnamefont#1{#1}\fi
\expandafter\ifx\csname bibfnamefont\endcsname\relax
  \def\bibfnamefont#1{#1}\fi
\expandafter\ifx\csname citenamefont\endcsname\relax
  \def\citenamefont#1{#1}\fi
\expandafter\ifx\csname url\endcsname\relax
  \def\url#1{\texttt{#1}}\fi
\expandafter\ifx\csname urlprefix\endcsname\relax\def\urlprefix{URL }\fi
\providecommand{\bibinfo}[2]{#2}
\providecommand{\eprint}[2][]{\url{#2}}

\bibitem[{\citenamefont{Sumikura et~al.}(2007)\citenamefont{Sumikura,
  Nagashima, Kitahara, and Hangyo}}]{hangyo}
\bibinfo{author}{\bibfnamefont{H.}~\bibnamefont{Sumikura} {\it et al.}},
%  \bibinfo{author}{\bibfnamefont{T.}~\bibnamefont{Nagashima}},
%  \bibinfo{author}{\bibfnamefont{H.}~\bibnamefont{Kitahara}}, \bibnamefont{and}
%  \bibinfo{author}{\bibfnamefont{M.}~\bibnamefont{Hangyo}},
  \bibinfo{journal}{Jpn. J. Appl. Phys.} \textbf{\bibinfo{volume}{46}},
  \bibinfo{pages}{1739} (\bibinfo{year}{2007}).

\bibitem[{\citenamefont{Ikebe and Shimano}(2008)}]{ikebe2008cds}
\bibinfo{author}{\bibfnamefont{Y.}~\bibnamefont{Ikebe}} \bibnamefont{and}
  \bibinfo{author}{\bibfnamefont{R.}~\bibnamefont{Shimano}},
  \bibinfo{journal}{Appl. Phys. Lett.} \textbf{\bibinfo{volume}{92}},
  \bibinfo{pages}{012111} (\bibinfo{year}{2008}).
  
\bibitem[{\citenamefont{Thouless et~al.}(1982)\citenamefont{Thouless, Kohmoto,
  Nightingale, and den Nijs}}]{tknn}
\bibinfo{author}{\bibfnamefont{D.~J.} \bibnamefont{Thouless} {\it et al.}},
%  \bibinfo{author}{\bibfnamefont{M.}~\bibnamefont{Kohmoto}},
%  \bibinfo{author}{\bibfnamefont{M.~P.} \bibnamefont{Nightingale}},
%s  \bibnamefont{and} \bibinfo{author}{\bibfnamefont{M.}~\bibnamefont{den Nijs}},
  \bibinfo{journal}{Phys. Rev. Lett.} \textbf{\bibinfo{volume}{49}},
  \bibinfo{pages}{405} (\bibinfo{year}{1982}).

\bibitem[{\citenamefont{Hatsugai}(1993)}]{hatsugai1993cna}
\bibinfo{author}{\bibfnamefont{Y.}~\bibnamefont{Hatsugai}},
  \bibinfo{journal}{Phys. Rev. Lett.} \textbf{\bibinfo{volume}{71}},
  \bibinfo{pages}{3697} (\bibinfo{year}{1993}).

\bibitem[{\citenamefont{Novoselov et~al.}(2005)\citenamefont{Novoselov, Geim,
  Morozov, Jiang, Katsnelson, Grigorieva, Dubonos, and Firsov}}]{Nov05}
\bibinfo{author}{\bibfnamefont{K.}~\bibnamefont{Novoselov} {\it et al.}},
%  \bibinfo{author}{\bibfnamefont{A.}~\bibnamefont{Geim}},
%  \bibinfo{author}{\bibfnamefont{S.}~\bibnamefont{Morozov}},
%  \bibinfo{author}{\bibfnamefont{D.}~\bibnamefont{Jiang}},
%  \bibinfo{author}{\bibfnamefont{M.}~\bibnamefont{Katsnelson}},
%  \bibinfo{author}{\bibfnamefont{I.}~\bibnamefont{Grigorieva}},
%  \bibinfo{author}{\bibfnamefont{S.}~\bibnamefont{Dubonos}}, \bibnamefont{and}
%  \bibinfo{author}{\bibfnamefont{A.}~\bibnamefont{Firsov}},
  \bibinfo{journal}{Nature} \textbf{\bibinfo{volume}{438}},
  \bibinfo{pages}{197} (\bibinfo{year}{2005}).

\bibitem[{\citenamefont{Zhang et~al.}(2005)\citenamefont{Zhang, Tan, Stormer,
  and Kim}}]{Kim-gr}
\bibinfo{author}{\bibfnamefont{Y.}~\bibnamefont{Zhang} {\it et al.}},
%  \bibinfo{author}{\bibfnamefont{Y.}~\bibnamefont{Tan}},
%  \bibinfo{author}{\bibfnamefont{H.}~\bibnamefont{Stormer}}, \bibnamefont{and}
%  \bibinfo{author}{\bibfnamefont{P.}~\bibnamefont{Kim}},
  \bibinfo{journal}{Nature} \textbf{\bibinfo{volume}{438}},
  \bibinfo{pages}{201} (\bibinfo{year}{2005}).

\bibitem[{\citenamefont{Sadowski et~al.}(2006)\citenamefont{Sadowski, Martinez,
  Potemski, Berger, and de~Heer}}]{sadowski}
\bibinfo{author}{\bibfnamefont{M.~L.} \bibnamefont{Sadowski} {\it et al.}},
%  \bibinfo{author}{\bibfnamefont{G.}~\bibnamefont{Martinez}},
%  \bibinfo{author}{\bibfnamefont{M.}~\bibnamefont{Potemski}},
%  \bibinfo{author}{\bibfnamefont{C.}~\bibnamefont{Berger}}, \bibnamefont{and}
%  \bibinfo{author}{\bibfnamefont{W.~A.} \bibnamefont{de~Heer}},
  \bibinfo{journal}{Phys. Rev. Lett.} \textbf{\bibinfo{volume}{97}},
  \bibinfo{pages}{266405} (\bibinfo{year}{2006}).

\bibitem[{\citenamefont{Morimoto et~al.}(2008)\citenamefont{Morimoto, Hatsugai,
  and Aoki}}]{morimoto-CE}
\bibinfo{author}{\bibfnamefont{T.}~\bibnamefont{Morimoto}},
  \bibinfo{author}{\bibfnamefont{Y.}~\bibnamefont{Hatsugai}}, \bibnamefont{and}
  \bibinfo{author}{\bibfnamefont{H.}~\bibnamefont{Aoki}},
  \bibinfo{journal}{Phys. Rev. B} \textbf{\bibinfo{volume}{78}},
  \bibinfo{eid}{073406} (\bibinfo{year}{2008}).

\bibitem[{\citenamefont{Nomura et~al.}(2008)\citenamefont{Nomura, Ryu, Koshino,
  Mudry, and Furusaki}}]{nomura2008qhe}
\bibinfo{author}{\bibfnamefont{K.}~\bibnamefont{Nomura} {\it et al.}},
%  \bibinfo{author}{\bibfnamefont{S.}~\bibnamefont{Ryu}},
%  \bibinfo{author}{\bibfnamefont{M.}~\bibnamefont{Koshino}},
%  \bibinfo{author}{\bibfnamefont{C.}~\bibnamefont{Mudry}}, \bibnamefont{and}
%  \bibinfo{author}{\bibfnamefont{A.}~\bibnamefont{Furusaki}},
  \bibinfo{journal}{Phys. Rev. Lett.} \textbf{\bibinfo{volume}{100}},
  \bibinfo{pages}{246806} (\bibinfo{year}{2008}).

\bibitem[{\citenamefont{Guinea et~al.}(2008)\citenamefont{Guinea, Horovitz, and
  LeDoussal}}]{guinea-ripple}
\bibinfo{author}{\bibfnamefont{F.}~\bibnamefont{Guinea}},
  \bibinfo{author}{\bibfnamefont{B.}~\bibnamefont{Horovitz}}, \bibnamefont{and}
  \bibinfo{author}{\bibfnamefont{P.}~\bibnamefont{LeDoussal}},
  \bibinfo{journal}{Phys. Rev. B} \textbf{\bibinfo{volume}{77}},
  \bibinfo{eid}{205421} (\bibinfo{year}{2008}).

\bibitem[{\citenamefont{Kawarabayashi et~al.}(2009)\citenamefont{Kawarabayashi,
  Hatsugai, and Aoki}}]{kawarabayashi09}
\bibinfo{author}{\bibfnamefont{T.}~\bibnamefont{Kawarabayashi}},
  \bibinfo{author}{\bibfnamefont{Y.}~\bibnamefont{Hatsugai}}, \bibnamefont{and}
  \bibinfo{author}{\bibfnamefont{H.}~\bibnamefont{Aoki}},
  \bibinfo{journal}{Phys. Rev. Lett.} \textbf{\bibinfo{volume}{103}},
  \bibinfo{pages}{156804} (\bibinfo{year}{2009}).

\bibitem[{\citenamefont{Ando}(1975)}]{ando}
\bibinfo{author}{\bibfnamefont{T.}~\bibnamefont{Ando}}, \bibinfo{journal}{J.
  Phys. Soc. Jpn.} \textbf{\bibinfo{volume}{38}}, \bibinfo{pages}{989}
  (\bibinfo{year}{1975}).

\bibitem[{\citenamefont{O'Connell and Wallace}(1982)}]{oconnell82}
\bibinfo{author}{\bibfnamefont{R.~F.} \bibnamefont{O'Connell}}
  \bibnamefont{and} \bibinfo{author}{\bibfnamefont{G.}~\bibnamefont{Wallace}},
  \bibinfo{journal}{Phys. Rev. B} \textbf{\bibinfo{volume}{26}},
  \bibinfo{pages}{2231} (\bibinfo{year}{1982}).

\bibitem[{\citenamefont{Peng et~al.}(1991)\citenamefont{Peng, Zhou, and
  Shen}}]{peng1991frq}
V. Volkov and S. Mikhailov, JETP. Lett. {\bf 41}, (1985);
\bibinfo{author}{\bibfnamefont{J.~P.} \bibnamefont{Peng}},
  \bibinfo{author}{\bibfnamefont{S.~X.} \bibnamefont{Zhou}}, \bibnamefont{and}
  \bibinfo{author}{\bibfnamefont{X.~C.} \bibnamefont{Shen}},
  \bibinfo{journal}{Phys. Rev. B} \textbf{\bibinfo{volume}{44}},
  \bibinfo{pages}{4021} (\bibinfo{year}{1991}).
 
\bibitem[{\citenamefont{Nair et~al.}(2008)\citenamefont{Nair, Blake,
  Grigorenko, Novoselov, Booth, Stauber, Peres, and Geim}}]{nair-alpha}
\bibinfo{author}{\bibfnamefont{R.}~\bibnamefont{Nair} {\it et al.}},
%  \bibinfo{author}{\bibfnamefont{P.}~\bibnamefont{Blake}},
%  \bibinfo{author}{\bibfnamefont{A.}~\bibnamefont{Grigorenko}},
%  \bibinfo{author}{\bibfnamefont{K.}~\bibnamefont{Novoselov}},
%  \bibinfo{author}{\bibfnamefont{T.}~\bibnamefont{Booth}},
%  \bibinfo{author}{\bibfnamefont{T.}~\bibnamefont{Stauber}},
%  \bibinfo{author}{\bibfnamefont{N.}~\bibnamefont{Peres}}, \bibnamefont{and}
%  \bibinfo{author}{\bibfnamefont{A.}~\bibnamefont{Geim}},
  \bibinfo{journal}{Science} \textbf{\bibinfo{volume}{320}},
  \bibinfo{pages}{1308} (\bibinfo{year}{2008}).

\bibitem[{\citenamefont{Morimoto et~al.}(2009)\citenamefont{Morimoto, Hatsugai,
  and Aoki}}]{morimoto-opthall}
\bibinfo{author}{\bibfnamefont{T.}~\bibnamefont{Morimoto}},
  \bibinfo{author}{\bibfnamefont{Y.}~\bibnamefont{Hatsugai}}, \bibnamefont{and}
  \bibinfo{author}{\bibfnamefont{H.}~\bibnamefont{Aoki}},  
  \bibinfo{journal}{Phys. Rev. Lett.} \textbf{\bibinfo{volume}{103}},
  \bibinfo{pages}{116803} (\bibinfo{year}{2009}).

\bibitem[{\citenamefont{Morimoto et~al.}(2009)\citenamefont{Morimoto, Hatsugai,
  and Aoki}}]{morimoto-LT-opthall}
\bibinfo{author}{\bibfnamefont{T.}~\bibnamefont{Morimoto}},
  \bibinfo{author}{\bibfnamefont{Y.}~\bibnamefont{Hatsugai}}, \bibnamefont{and}
  \bibinfo{author}{\bibfnamefont{H.}~\bibnamefont{Aoki}}, \bibinfo{journal}{J.
  Phys.: Conf. Ser.} \textbf{\bibinfo{volume}{150}}, \bibinfo{pages}{022060}
  (\bibinfo{year}{2009}).

\bibitem[{\citenamefont{Aoki and Ando}(1981)}]{aoki-ando1981}
\bibinfo{author}{\bibfnamefont{H.}~\bibnamefont{Aoki}} \bibnamefont{and}
  \bibinfo{author}{\bibfnamefont{T.}~\bibnamefont{Ando}},
  \bibinfo{journal}{Solid State Commun.} \textbf{\bibinfo{volume}{38}},
  \bibinfo{pages}{1079} (\bibinfo{year}{1981}).

\bibitem{aoki78} H. Aoki, J. Phys. C {\bf 11}, 3823 (1978).

\bibitem[{\citenamefont{Nomura et~al.}(2007)\citenamefont{Nomura, Koshino, and
  Ryu}}]{nomura-ryu-koshino}
\bibinfo{author}{\bibfnamefont{K.}~\bibnamefont{Nomura}},
  \bibinfo{author}{\bibfnamefont{M.}~\bibnamefont{Koshino}}, \bibnamefont{and}
  \bibinfo{author}{\bibfnamefont{S.}~\bibnamefont{Ryu}},
  \bibinfo{journal}{Phys. Rev. Lett.} \textbf{\bibinfo{volume}{99}},
  \bibinfo{pages}{146806} (\bibinfo{year}{2007}).

\bibitem[{\citenamefont{Zheng and Ando}(2002)}]{ZhengAndo}
\bibinfo{author}{\bibfnamefont{Y.}~\bibnamefont{Zheng}} \bibnamefont{and}
  \bibinfo{author}{\bibfnamefont{T.}~\bibnamefont{Ando}},
  \bibinfo{journal}{Phys. Rev. B} \textbf{\bibinfo{volume}{65}},
  \bibinfo{pages}{245420} (\bibinfo{year}{2002}).


\end{thebibliography}
\end{document}